# HOPES - An Integrative Digital Phenotyping Platform for Data Collection, Monitoring and Machine Learning

Xuancong Wang[1], Nikola Vouk[1], Creighton Heaukulani[1], Thisum Buddhika[1], Wijaya Martanto[1], Jimmy Lee[2,4], Robert JT Morris[1,3]

[1]Ministry of Health Office for Healthcare Transformation (MOHT), Singapore
[2]Institute of Mental Health, Singapore
[3]Yong Loo Lin School of Medicine, National University of Singapore, Singapore
[4]Lee Kong Chian School of Medicine, Nanyang Technological University, Singapore

Corresponding Author:     Creighton Heaukulani (creighton.heaukulani@moht.com.sg)

Abstract

We describe the development of, and early experiences with, a comprehensive Digital Phenotyping platform: **H**ealth **O**utcomes through **P**ositive **E**ngagement and **S**elf-Empowerment (HOPES). HOPES is based on the open-source *Beiwe* platform but adds a much wider range of data collection, including the integration of wearable devices and further sensor collection from the smartphone. Requirements were in part derived from a concurrent clinical trial for schizophrenia. This trial required development of significant capabilities in HOPES for security, privacy, ease-of-use and scalability, based on a careful combination of public cloud and on-premises operation. We describe new data pipelines to clean, process, present and analyze data. This includes a set of dashboards customized to the needs of research study operations, and for clinical care. A test use case for HOPES is described by analyzing the digital behavior of 20 participants during the SARS-CoV-2 pandemic.

## 1 Introduction

We are at an age in healthcare where we have much data at our disposal, including high penetration of digital electronic medical records and advanced techniques available for their analysis [1]. It is also well-accepted that the characterization of lifestyle, including activity, stress level, social interactions, and environment, are significant determinants of health outcomes [2,3]. While estimates vary, it has been argued that "lifestyle choices" exceed the impact of "healthcare received" as a determinant of premature death [3].

The wide adoption of smartphones and the increasing use of wearable devices opens up a new vista of characterizing both current health conditions and the ongoing behaviors that will determine how an individual's health will evolve. As examples of these new data sources, we can readily measure physical activity, heart rate, heart rate variability, temperature, sleep, "sociability" (amount of human interaction), smartphone usage (amount and duration of use, type of use, the way a screen is tapped and scrolled), etc. The approach of using personal digital devices to capture these data sources, and hence characterize an individual *in situ,* has been called *digital phenotyping* [4]. The use of digital phenotyping both complements and extends the use of traditional home monitoring (e.g., blood pressure measurements) in telemedicine by offering continuous measurement during normal activities and everyday living. We have developed a general-purpose digital phenotyping platform **H**ealth **O**utcomes through **P**ositive **E**ngagement and **S**elf-Empowerment (HOPES), that integrates data from wearable devices and a broad set of smartphone sensors, provides an array of methods to inspect that data, and binds everything together into a platform with a comprehensive privacy and security model. The platform was developed in conjunction with the running of a clinical study for Schizophrenia. In what follows, we will overview the HOPES platform and demonstrate its first use.

## 1.1 Digital Phenotyping

The collection of data from a personal digital device can be used to encourage healthy behaviors, for example the Singapore Health Promotion Board National Steps Challenge [5]. Data collection is sometimes combined with coaching or nudges for general wellness [6], or to monitor and improve an existing diagnosed condition [7,8]. We were originally inspired by the potential use of digital phenotyping to monitor and treat *mental health* conditions such as depression and schizophrenia. Several notable studies include a study at Northwestern University that showed the correlation of mobile phone sensors with depressive symptom severity [9], a recent study at King's College London which showed the feasibility and acceptability of the extended use of wearable devices and smartphones in schizophrenia patients [10], and the use of digital phenotyping for relapse prediction in schizophrenia [11]. The commercial world has also taken notice and a number of start-ups have formed [12,13,14].

In addition, digital phenotyping is also being applied to address a diverse range of diseases such as asthma, maternal health, cancer and dermatology [8,15,16,17]. Recent experiments in the use of wearable devices such as the Oura ring [18] and the Fitbit wristband [19] are being applied to measure participants' parameters during the SARS-CoV-2 pandemic. We will illustrate our observations related to the pandemic (using Fitbit) later in Section 5.

Digital phenotyping has the potential to supplement (and in some cases replace) standard clinical processes in data gathering and patient monitoring by virtue of the following attributes:

a. **Productivity and cost**: passive monitoring can be efficient for both the provider and patient compared traditional clinical visits or scheduled telemedicine encounters.

b. **Latency**: passive monitoring may enable relatively quick responses from health providers, e.g., allowing for actions from a case manager within a day versus a week or longer for a visit.

c. **Sensitivity**: several variables such as resting heart rate or sleep parameters are not easy to measure in the clinic, and device monitoring can be more effective than subjective patient reporting or inconvenient manual processes. The emergence of low-cost consumer devices has been shown to be accurate enough for several of these purposes [20].

d. **More parameters**: while in the past we have been limited to infrequent interview questions and scales, we now have the potential of monitoring a wider variety of the subjects' parameters, such as location, sleep, motion, heart rate variables, instantaneous manual responses, etc. These can be measured simultaneously, efficiently and reliably.

A much-discussed concern is how well such techniques will be accepted, and complied with, by patients or consumer participants. This involves ease-of-use considerations by both user and provider. Another major concern revolves around data security and privacy preservation. These two aspects have been primary motivators in our design choices and investigations.

## 1.2 Clinical Study on Digital Phenotyping in Schizophrenia

The HOPES platform was designed, developed and refined concurrently to support a clinical study. The HOPE-S (**H**ealth **O**utcomes via **P**ositive **E**ngagement in **S**chizophrenia) study [21] was launched in November 2019. HOPE-S is an observational study on individuals with schizophrenia who were recently discharged from a psychiatric hospitalization. The aim of the study is to determine whether digital phenotyping data is associated with clinical and health utilization outcomes. Key events recorded over the 6-month observation period include readmission, outpatient non-attendances (i.e., defaults) and unscheduled service use, e.g., emergency department attendance and Mobile Crisis Team (MCT) activations. The primary study outcomes are the ability to predict relapse and/or readmission within 6 months, with secondary outcomes being the associations between digital phenotyping data and healthcare utilization, psychiatric symptoms severity and functional status assessed during research visits. Ethics approval has been granted by Singapore's National Healthcare Group (NHG) Domain Specific Review Board (DSRB Reference no.: 2019/00720) and it has been registered on clinicaltrials.gov.

The first phase of HOPE-S is observational. During this phase, we are examining the deployment feasibility and acceptability of a wide range of digital sensors, while performing the analyses required to achieve the outcomes illustrated above. In the process we are collecting large amounts of data for our analyses. This data will subsequently be used to develop machine learning algorithms to *predict* changes in symptom severity and other important clinical outcomes, as opposed to merely analyzing associations. During a subsequent phase of the study, we will deploy interventions such as early warnings of relapses, which will allow pre-emptive steps to be taken to prevent participant setbacks or rehospitalization.

## 1.3 HOPES: A General-Purpose Platform for Digital Phenotyping

HOPES stands for **H**ealth **O**utcomes through **P**ositive **E**ngagement and **S**elf-Empowerment. It is based on, and extends, the existing platform *Beiwe* [22,23]. Our contributions include:

1. the integration of wearable devices, where we have experimented with both wrist and ring devices;
2. the use of further sensors on the smartphone;
3. an efficient onboarding method for participants;
4. a suite of user interfaces including data collection and quality management tools, clinical summarization dashboards, and general-purpose research dashboards for use in exploratory data analysis and building anomaly detection algorithms;
5. assurances for data security and the preservation of user privacy.

The platform is designed to be reliably deployed at scale and makes use of both public cloud and controlled on-premises computing infrastructure. We recognize the broad spectrum of potential applications beyond mental health and the growing set of digital sensors and their capabilities that may be appropriate for different applications. We have therefore designed HOPES to be flexible and extensible to accommodate new device and sensor integration, new data dashboards, etc.

While the data collected during the HOPE-S study is rich, it is also at times noisy and incomplete, as is to be expected when dealing with real human behavior and varying data reliability among sensors. To address these challenges, we have developed a *data collection dashboard* and multiple data visualization and exploration tools, which have proven invaluable for monitoring and ensuring participant compliance on a daily basis in the research study. We have also developed a feature engineering pipeline to construct useful insights for the HOPE-S study and to compensate for various shortfalls in the raw data. These dashboards have been found to be easy to use by research coordinators involved in the HOPE-S study, who have been able to easily recognize problems and contact the participant if their data is not being received. We will also illustrate the dashboards that our data scientists have used to look for patterns and an *anomaly detection dashboard* that raises alerts on irregularities in the data. All the data then feeds into full-blown statistical analysis and our ongoing development of predictive machine learning algorithms.

The rest of the paper is organized as follows. In Section 2, we review several existing open-source digital phenotyping platforms, highlighting their respective strengths and weaknesses. In Section 3, we describe the overall architecture of the HOPES platform. Section 4 describes the enhancements to *Beiwe* that the HOPES platform provides, guided by the requirements of the HOPE-S study and other planned future uses (including for purposes beyond mental health). In Section 5, we show an early and simple example of the use of our collected data on 20 participants in which we compare user data before and after Singapore's SARS-CoV-2 "lockdown" went into effect. In Section 6, we give some overall conclusions we can draw from our experiences in digital phenotyping.

## 2 Existing Platforms

There are several existing open-source digital phenotyping platforms, including *Beiwe* [22,23], *Purple Robot* [24,25,26], *AWARE* [27,28], and *RADAR-base* (Remote Assessment of Disease And Relapse) [29,30,31]. Each contains a core smartphone application (or "app") that performs passive sensor data collection in the background and a server backend in charge of receiving the data. Note that digital phenotyping is not limited to smartphones – indeed wearables also provide some significant differentiated capabilities and there are other sources such as fixed detectors. Some platforms such as *Beiwe* and *RADAR-base* support active data collection in the form of surveys and/or capture data from wearable devices such as wrist- or arm-wearable devices by providing a common data interface.

From our assessment, Purple Robot has the most complete coverage of Android sensors and features amongst the platforms we reviewed. The user can select which sensors to turn on and set the data sampling frequency, however, the platform does not support iOS. AWARE supports both Android and iOS, and has nearly full coverage of Android sensors and features. Like Purple Robot, it also allows the user to configure sensors and features. RADAR-base has recently added iOS support and uses both passive (phone/sensor) and active (survey and questionnaire) data collection. Although it covers fewer phone features and sensors than Purple Robot and AWARE, it has a very attractive user interface and very robust system for surveys and questionnaires. *Beiwe* is a smartphone-based digital phenotyping research platform that supports both Android and iOS, and has a decent coverage of phone sensors and features. Moreover, the platform supports active feature collection from simple surveys. Apart from the data collection backend that receives data from participants' phones, *Beiwe* also has a backend for data analytics.

We have based the framework for the HOPES platform on *Beiwe* for several reasons. Firstly, *Beiwe* supports both Android and iOS, a requirement for any generic digital phenotyping platform to be widely adopted. Secondly, our platform analysis and comparison tests during

March 2019 showed that *Beiwe* was the most ready at that time to deploy. Our decision was also based on our review of a number of Git repositories and publications, as well as previous practical applications of the platforms in clinical studies and trials.

To access data beyond the smartphone sensors, we choose the Fitbit wrist device. after carrying out a technical and usability comparison between several popular devices in the commercial market. Specifically, we have compared Fitbit Charge 3, Huawei Honor A2, Xiaomi Mi Band 3, Actxa Spur+ and Hey Plus. We found that Fitbit was distinguished by ease-of-use, battery life, and reliability and it has been validated to be reasonably accurate against gold standard devices for measurement of sleep [20]. We also evaluated a number of external sleep measurement devices (such as mattress pads) but did not find them suitable for our purposes.

## 3 The HOPES Platform and Its First Use in the HOPE-S Study

To support large scale data aggregation of wearables, mobile phones, and other data sources, we defined a set of requirements and then built our platform to be secure and scalable. Building on top of the existing *Beiwe* platform, we created the HOPES platform by expanding the functional capabilities for easier participant onboarding, enhanced data collection monitoring, optimized data upload, extended security features, expanded data processing and analytics pipeline, and a scalable deployment architecture. The goal was to obtain easy and secure onboarding, almost unlimited scaling, high operational security and improved privacy assurances. While we were immediately driven by meeting the strict requirements for the HOPE-S study, along the way we became aware of expanded requirements for a wider range of participant monitoring requirements and we took those into account in our architecture and design, so we would be ready for further deployments. In this section we describe the platform requirements and our resulting HOPES system architecture; we then detail the features collected for the HOPE-S study, the enhancements we made to the Android app, the platform backend and the security protocols. We provide our motivation and a high-level description, leaving further details and information about miscellaneous improvements to the supplementary materials.

### 3.1 Platform Requirements

The HOPES platform is designed to be a reliable, low-maintenance digital phenotyping collection and aggregation platform. It is designed to support research protocols as well as scale to larger production platforms including self-service registration. The requirements and their implemented capabilities are enumerated in Table 1.

Table 1 HOPES Platform Requirements

| Requirements | Implementation Capabilities |
|---|---|
| Simple user onboarding | <ul><li>Pre-creation of user identities and anonymization factors</li><li>Pre-printed QR code onboarding sheets</li><li>Ability to migrate participants to new phones (if their current phones are not usable for a study) while maintaining study data integrity and privacy</li><li>Simple Onboarding Literature and packaging in "gift pack" format</li><li>Wide Platform Support, Android, iOS, post-Android Chinese phones</li><li>Ability to be totally passive and with zero user interaction after setup</li><li>Preparation for self-service onboarding in the Future</li></ul> |
| User data collection and privacy | <ul><li>All data de-identified (zero PII)</li><li>Per-participant encryption keys</li><li>Per-participant random credentials</li></ul> |
| | • Mapping between patient/participant ID and de-identified Study ID securely retained but only made available to authorized clinicians<br>• Secure data backup and archiving |
| Data Security is end-to-end | • Data encrypted while in cloud storage environment<br>• Data decrypted but still deidentified, and obfuscated where appropriate, in data analytics pipeline on-premise |
| Wearable support must be scalable and secure | • Pre-creation of wearable device accounts<br>• Wearable cloud accounts deidentified using Study IDs<br>• Wearable data automatically encrypted with user's password<br>• Serverless functions to periodically collect and archive user data |
| Infrastructure, scale, and operational security | • Two-Factor authentication for all participants, including certificate and credential authentication<br>• Rotating Credentials<br>• Data Collection Dashboard<br>• HOPES Work/Ticket Queue for monitoring Alerts/Logs/Events<br>• DDoS/Web Application Firewall protection<br>• Elastic Scale at all levels<br>• Isolation of functions across private Virtual Private LAN<br>• Separation of Administrative and Data Upload Interfaces<br>• Private VPN for Administration<br>• Separation of Data Upload API and Data Management<br>• Restricted Access controls<br>• Automated Repeatable Deployment |
| Data Analytics | • Data downloaded on-premise into secure workspace for analytics or clinical use<br>• Multi-stage analytics processing pipeline<br>• Anomaly Detection Dashboard<br>• Data Exploration Dashboard |
| Security Standards | • Secure Development Process<br>• Automated Patching<br>• Additional requirements from Singapore Security and IT standards |
| Expanded Data Collection Support for Social Media Metadata | • Support for de-identified metadata for WhatsApp text and audio/video messages on Android<br>• Phone Text Messages |
| Study Clinical Support for easy Clinical Management | • Daily De-identified Data Collection Dashboard emailed to study researchers and clinicians to monitor study compliance<br>• Encrypted Deidentified Clinician Dashboard accessible to clinicians |

To successfully implement such a broad set of requirements, we carefully studied and focused on the user experience for onboarding new participants, and built a platform that leverages the best software engineering, design principles, and cloud architecture capabilities.

## 3.2 Overall System Architecture

The high-level solution architecture of HOPES, as used for the HOPE-S study, is shown in Figure 1. On each participant's smartphone, we install two apps: the Fitbit app and the HOPES app (on the smartphone). Every participant is required to wear a Fitbit watch for a certain portion of the day (enough to collect required data, but also removable for charging, showering, etc.). Fitbit raw data is collected by the Fitbit app and sent to a Fitbit server (the *Fitbit Cloud*) for processing and computation of high-level features (e.g., the estimation of sleep stages). Phone data is collected on the smartphone by the HOPES app and sent to *data upload server* hosted in a public cloud - we used Amazon Web Services (AWS). The *data processing backend server* (at the R&D or clinical premises) periodically pulls data from both the Fitbit cloud and the AWS *data upload server* for subsequent processing, described in the following sections. The data is always de-identified when in a publicly accessible cloud environment, and all transmission and storage are encrypted. Certain variables such as location are also obfuscated at time of collection for privacy preservation. More details on the solution architecture are described in the supplementary material.

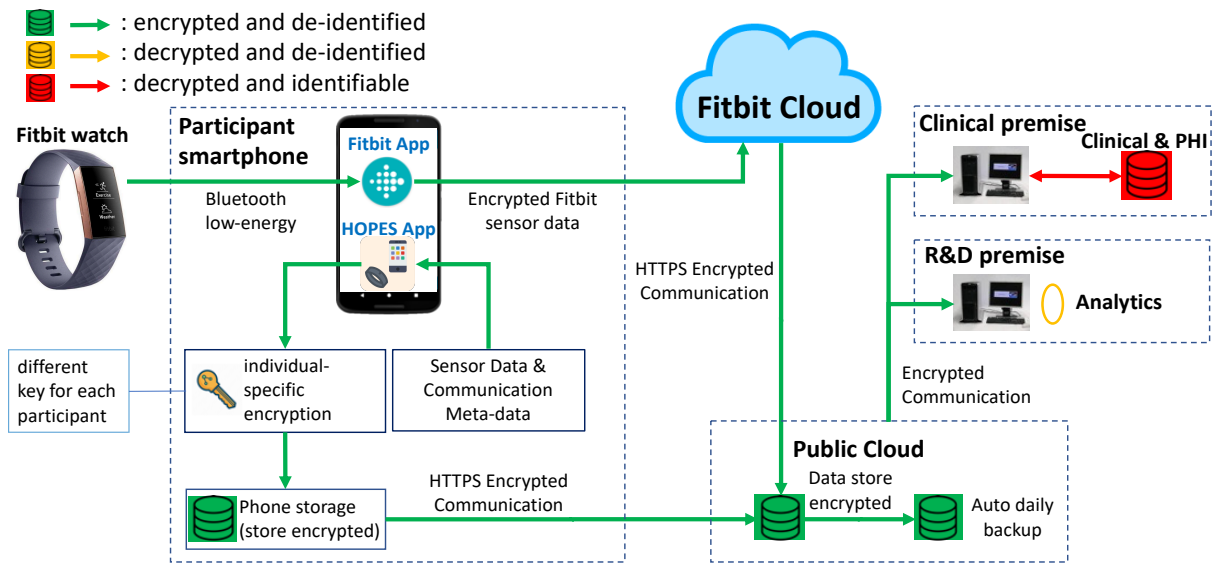

Figure 1: Overall system architecture and data-flow diagram for the HOPE-S Study

For the backend R&D analytics we developed a set of data processing pipelines and various dashboards for monitoring, visualizing and analyzing data (as shown in Figure 2). The data processing pipelines clean (manage missing, duplicated and erroneous data), convert and re-organize data into more usable forms. These dashboards are used by research coordinators and clinicians, researchers, data analysts and technical team members involved in the conduct of the study. A general-purpose research dashboard supports exploratory analytics. In each case, roles and responsibilities determine access controls to various attributes of the data. Physical controls, supervision and accountability measures are also deployed to make sure there is no unauthorized access to data. Further description is given in subsequent sections, and greater detail in the supplementary material.

### 3.3 Features Supported by the Platform

The following six categories of features are obtained from the HOPES smartphone app. In each case we will indicate "new" if it is a new feature added by us or an enhancement, otherwise, it is an existing feature in the *Beiwe* distribution:

**Location:** GPS coordinates are used to detect deviations from typical travel patterns and to compute a measure of variance or entropy in the locations visited by a participant. To protect user privacy, the raw GPS coordinates are obfuscated via a random displacement (from the origin) which is unique for every participant.

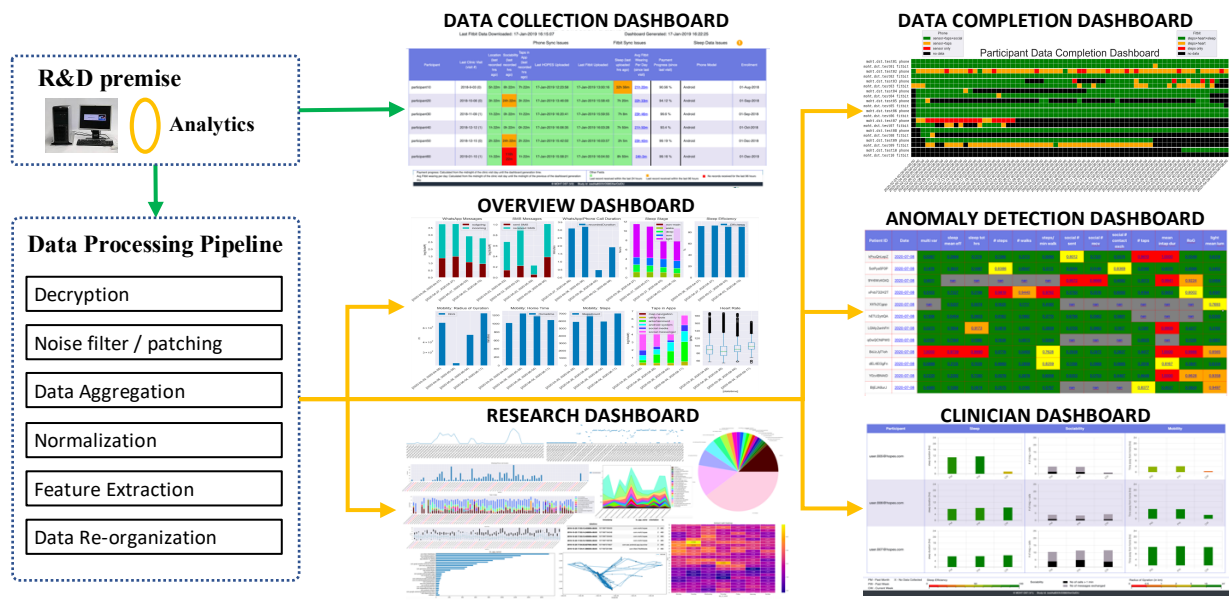

Figure 2: Backend data processing pipelines and dashboards

**Sociability indices (some are new):** For our study, changes in a participant's s*ociability*, i.e., their communication with others, is estimated from available data. Sociability may be reflected in their activity in various forms of messaging and voice/video communications. The original *Beiwe* app can capture incoming and outgoing phone calls and SMS messages. However, in many countries most people use free social messaging apps as their primary method for text and voice communication, e.g., in Singapore WhatsApp is dominant. We therefore make use of the Android Accessibility Service API to acquire message metadata from social messaging apps. So far, we have only implemented this for WhatsApp but it can be easily extended to other social messaging apps. The duration and timing of mobile service phone calls and WhatsApp calls sent and received are recorded. Likewise, the length and timing of SMS and WhatsApp messages sent and received are also recorded. Importantly, for privacy protection, we never record or transmit any content of any communication, and we hash the identity or contact number of the counterparty.

**Finger taps (new):** Taps provide two types of information that may be related to a person's health. The speed at which a person taps may give a hint of their neuropsychological function [32]; for example, a fatigued person may tap more slowly, or some diseases may cause small, uncontrollable movements. There is also some evidence that finger taps may be used to detect depression [33]. The apps a person uses (determined from their taps) also gives an indication of their behavior. For example, a patient with mental illness who is relapsing might be found to have significantly altered communications, reflected in the number and speed of taps made in the various apps. We seek to capture typing error rates, which could be affected by physical or mental condition. We can determine this from how often the delete or backspace key on the keyboard is tapped. To measure tapping speed, we also need to know whether the person is typing on the keyboard or navigating in a social messaging app. Characteristics and metadata of finger taps on the phone screen are recorded, such as the number and timestamps of taps into apps, different key strokes (from the enter key, delete key, backspace key, alphabet keys, number keys, and punctuation keys), and the group categorization of the apps tapped are also recorded. As a privacy preservation measure, captured keystrokes are converted into a token (such as "alphabetic", "numerical", "punctuation", etc.). The app will only store and download the token, and the specific keys that are struck are not recorded.

**Motion info (some are new):** Accelerometer, gyroscope, magnetometer (new), and pedometer (new) data are recorded to check whether the phone is being moved or motionless. This information can help determine the amount of phone use, and can be corelated with other data collected with wearables (sleep, activity, etc.).

**Phone states:** The app can record the Wi-Fi state, the Bluetooth state, and the power state (screen on/off and power-down event) of the phone. The Wi-Fi and Bluetooth scan result can to some extent tell some information about the location of the device especially when the GPS location is not available. However, this data is sensitive and needs to be de-identified and encrypted. The power state feature is usually combined with other features such as taps to tell the usage behavior of the phone by the participant.

**Ambient light (new):** The app can record the intensity of ambient light through the smartphone's built-in light sensor (not the camera). This could detect, for example, whether a participant is in a comfortable sleeping environment, and studies have suggested there is correlation between a patient's mental health and their preferred environmental lighting [34].

Since sleep and heart-rate are important indications of people's mental health status, we record the following three categories of features from the Fitbit wearable (obtained directly from the Fitbit cloud).

**Sleep:** Sleep information during the day and night are recorded, including a breakdown of different sleep stages with timestamps.

**Steps:** The total number of steps in time intervals specified by Fitbit.

**Heart rate:** The number of heart beats in time intervals specified by the Fitbit. Approximations of other measures of interest, such as heart rate variability, may be computed from the heart rate data.

For the HOPE-S study, we have used the following features from above: location, sociability indices, finger taps, accelerometer, power state, ambient light, sleep, steps, and heart rate.

### 3.4 Backend Data Processing Pipeline

We have rebuilt *Beiwe* data processing back-end in Python 3 to systematically process data files, reformat the raw data, and extract high-level features. A considerable amount of feature engineering is being performed on the backend to clean the data, correct data shortcomings, combine different data sources into joint features, and feed downstream machine learning systems. For example, upon consultation with our clinical partners, we construct high level features that are likely to provide useful signals regarding the mental health of the participants in the HOPE-S study. Our current analyses in the study make use of time series of daily or hourly samples of intuitively-identified measurements from sleep, steps, heart rate, location, and sociability indices. Some examples include daily totals of the number of hours of sleep, steps, and of communications initiated and received. Constructing such features is often necessary in situations with small amounts of or noisy data. As an example, when no sleep data is recorded by the Fitbit for a whole day, it isn't clear whether the participant didn't sleep, or whether they just didn't wear the Fitbit to bed. We can resolve this ambiguity by looking at the heart rate measurements, which are recorded continually while the Fitbit is worn. If heart rate data is missing for more than allotted allowance, we can reasonably assume the participant wasn't wearing the Fitbit during sleep. As another example, we have developed an Android app grouper that uses information from the Google Play Store to classify all apps into seven classes defined by us (i.e., social messenger, social media, entertainment, map navigation, utility tools, games, and android system (other vendor-specific or system apps that cannot be found in Play Store)). This class information is used in the taps data features when classifying a user's phone activity, e.g., "in social media apps", "in gaming apps", etc. In summary, this step bridges the gap between data collection and downstream machine learning modules. Details on the data processing pipeline, high-level feature extraction, and the seven classes of the app grouper are given in the supplementary material.

## 3.5 Platform Improvements

We have made many improvements to the Android app and are in the process of extending these improvements to the iOS app. In this section, we will only describe the most significant improvements, other improvements are in the supplementary material. We also use two system variations: the *prototype* or development system and the *deployed* system. Some features may apply to only one of the systems.

### 3.5.1 Scanning QR Codes for Simple User Registration

To facilitate the user registration process and to allow one-way encryption for better data security, study participant kits were prepared and a single page onboarding document was generated with all the information necessary to onboard a participant. The process is designed for a non-technical self-service onboarding process. In *the deployed system*, multiple QR codes are scanned. They include information for certificate-based authentication to further strengthen security via host verification. For details on QR registration, please refer to Section 2 of the supplementary material.

### 3.5.2 Data Compression

In order to scale up to a very large number of users, we need to reduce the utilized communication bandwidth as much as possible. One solution is to compress the data before sending it to the server. We have therefore added this option when creating a study, which may be selected on the backend console by checking the "enable compression" checkbox. Since the efficiency of compression is reduced significantly on encrypted data compression has to be done before encryption. This feature is implemented only in *the prototype system*.

### 3.5.3 Security Enhancements

The HOPES platform is re-designed on top of *Beiwe* to ensure data confidentiality, data integrity, system auditing, high availability, authentication, authorization to support large scale deployments with a distributed pipeline and separation of duties throughout the architecture to minimize the risk of data breach and preserve the privacy of data throughout the lifecycle. In the original *Beiwe* platform, data is decrypted in the data collection server and re-encrypted using the study key. This poses certain amount of risk because the data collection server is directly facing public Internet. In our HOPES platform, data is encrypted at all times while on the phone and in the data collection infrastructure, and is only decrypted in clinical or R&D premise. The decryption key is only accessible by clinical or R&D premise, so in principle the data is not decryptable on the phone nor in the data collection infrastructure. Data is only re-identified when needed for qualified clinical purposes, and only by clinical staff.

## 4  Dashboards, User Interfaces and Data Analysis

Ensuring complete data collection is important. A variety of issues can result in not receiving data as expected, including technical failures, participants not adhering to the guidelines on the device usage, or participants failing to wear their device. Monitoring this process becomes especially challenging at scale. We have therefore created a *data collection dashboard* (see Figure 3) to facilitate monitoring of data collected.

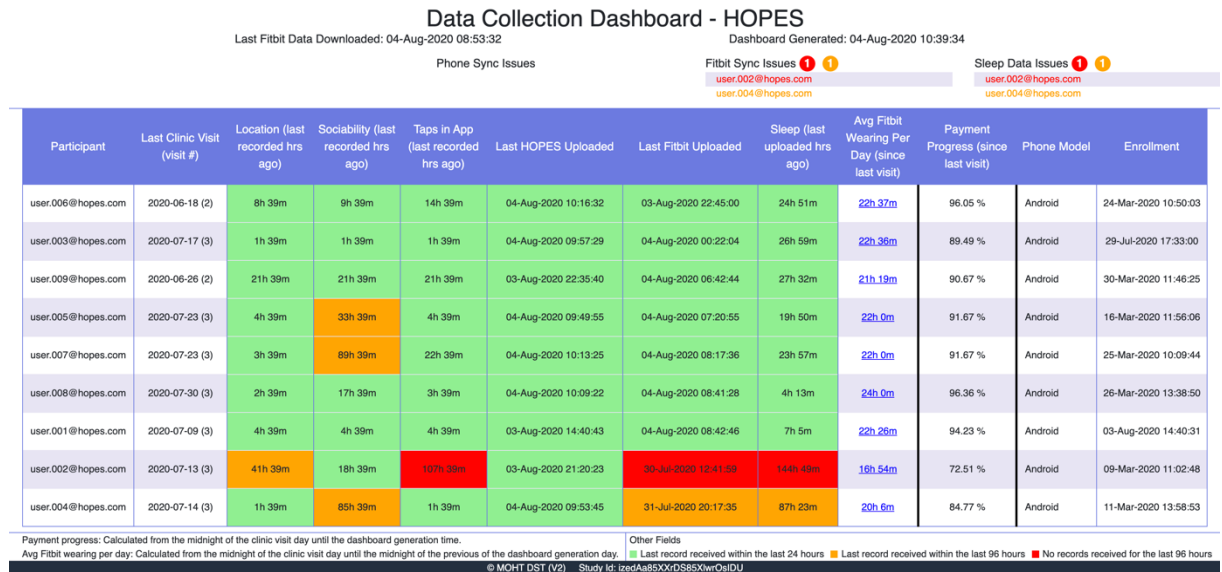

Figure 3: The data-collection dashboard shows the data uploading status of all participants.

*The data collection dashboard* is populated using the metadata extracted during the downloading phase of Fitbit and Phone data. The AWS Lambda function (which is set to trigger every five minutes) is set up to retrieve these data from their respective S3 buckets and creates an HTML file. To fill the dashboard so as to ensure that the participants comply with the study requirements, the following data types are observed and closely monitored: Location, Sociability, Taps in App, Last HOPES Uploaded, Last Fitbit Uploaded and Sleep. Color codes denote the severity of the data collection status, red being "need to take an action", orange means "need to closely monitor" and green being "normal".

The data collection dashboard does not require decrypted data and thus is constructed before decryption. As a result, it can be hosted on the upload server with little security risk. However, it does not show full historical data completion status which is sometimes needed. Hence, we developed the data completion dashboard which is described in detail in the supplementary material.

### 4.1  Data Visualization Toolkit

We have developed a *data visualization toolkit* for visualizing and exploring the collected data. The toolkit can also perform some basic statistical analyses, such as the comparison of features between defined date ranges. Figure 4 gives an illustration of the various types of graphs that can be plotted using the visualizer; these include most of the features discussed above and can be viewed in graphical form. For further details on the usage and capability of the data visualizer, please refer to the supplementary material.

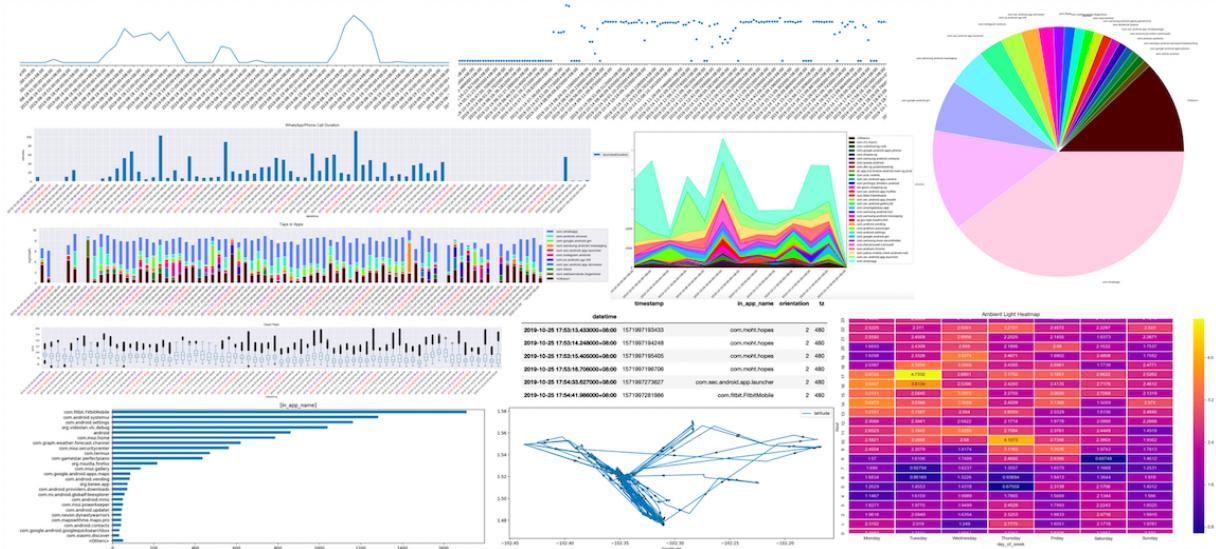

Figure 4: Types of graphs that can be plotted using the data visualizer.

## 4.2 Clinician Dashboard

The *clinician dashboard*, illustrated in Figure 5 is designed for clinicians to preview general trends in participants' digital marker data, and may be useful during clinical encounters. Based on previous studies and the observations of our clinical partners, we have decided to report sleep, sociability and mobility data for the current version of the clinician dashboard.

*Sleep* is plotted based on total sleep duration and sleep efficiency; the latter depicted by the color. *Sociability* is plotted using number of messages exchanged and the number of calls of duration more than one minute. *Mobility* is based on the *time away from home* (time spent away from sleeping location) and the *radius of gyration* (maximum distance travelled from home). These graphs are drawn based on averages over three time-frames: the current week is seven days before 0:00 am of the current day; the past week is 7 days prior to the current week; the past month is 30 days prior to the past week. An example further explaining the clinician dashboard can be found in the supplementary material.

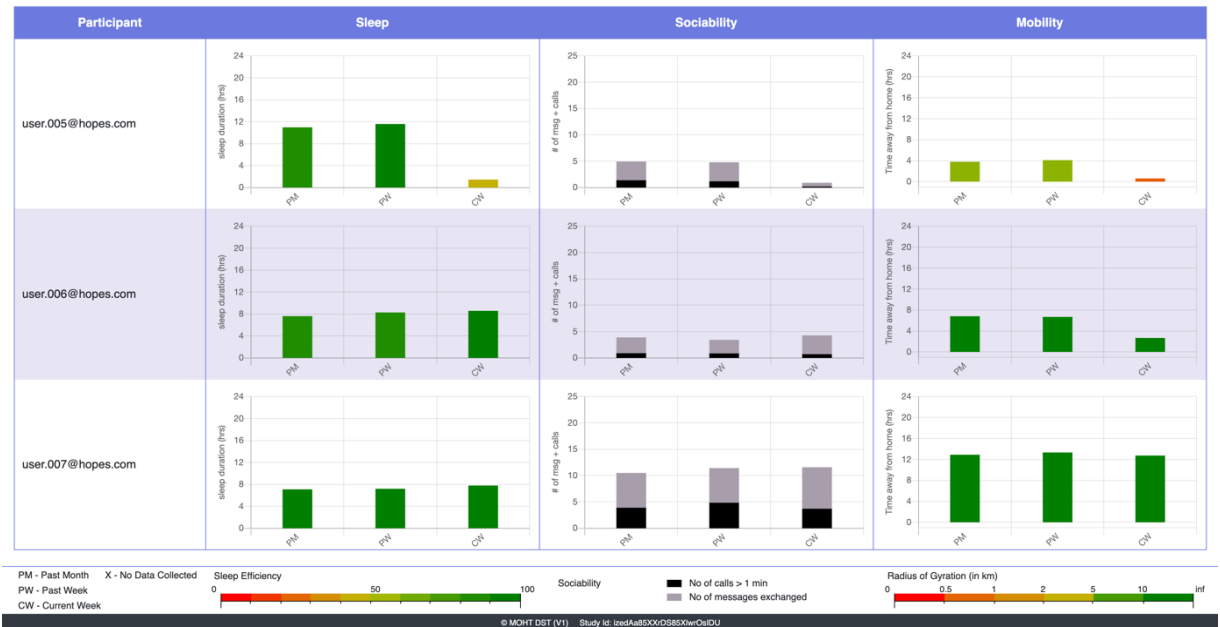

Figure 5: Clinician dashboard shows a preview of general trends in patient bio-marker data

## 4.3 Anomaly Detection Dashboard

In order to support a wide variety of applications attempting to analyze and identify interesting changes amongst the many features being collected by the platform, we have implemented a generic purpose *anomaly detection system and dashboard*. The system is comprised of several anomaly detection algorithms on the backend that report their findings via an *anomaly detection dashboard*. The dashboard is designed to create alerts about possible irregularities arising in the digital phenotyping data each day.

There are many approaches in machine learning to anomaly detection in time series data. One approach is to train a time series model on historical data from the participant and then have that model forecast what the data will look like in the future. When new data arrives, we compare it with the forecast and *score* the prediction based on how "good" or "bad" it is. For example, a simple scoring mechanism compares the empirical distribution of the *residuals* (i.e., the errors of the fitted model's predictions on the training set) to the realized prediction error on new data.

We have experimented with several time series models, including the broad class of autoregressive integrated moving average (ARIMA) models [35] and the class of *Gaussian processes* [36], fitting them to a subset of digital phenotyping features that we have initially selected as important for our HOPE-S study (see the supplementary material for details of the features). We note that these two choices of models are able to capture *periodic effects*, which are important for our HOPE-S study, since participants' behaviors may change markedly on the weekends. Selecting the most appropriate model will depend on the data and application at hand. We train the models every day on all past data and compute the predictions of the digital phenotyping features for the next day. At the end of the following day, the realized digital phenotyping features are compared to the predictions and scored, and these scores are transformed to be interpreted as "the probability that the observed data is an anomaly." The final score is therefore a number between zero and one, where higher values constitute alerts.

| Patient ID | Date | multi var | sleep mean eff | sleep tot hrs | # steps | # walks | steps/ min walk | social # sent | social # recv | social # contact exch | # taps | mean intap dur | RoG | light mean lum |
|---|---|---|---|---|---|---|---|---|---|---|---|---|---|---|
| kPxuQnLepZ | 2020-07-08 | 0.0267 | 0.2984 | 0.1219 | 0.1266 | 0.3173 | 0.5090 | 0.8012 | 0.7331 | 0.5574 | 0.9610 | 1.0000 | 0.5049 | 0.6319 |
| 5otPyaSF0P | 2020-07-08 | 0.1478 | 0.3631 | 0.1497 | 0.8386 | 0.6547 | 0.2172 | 0.5916 | 0.5148 | 0.8369 | 0.2140 | 0.4370 | 0.4588 | 0.3452 |
| 9Yr6WzKGtQ | 2020-07-08 | 0.8011 | nan | nan | nan | nan | nan | 0.9572 | 0.9666 | 0.4092 | 0.3573 | 0.9941 | 0.9224 | 0.0494 |
| xPvb732H2T | 2020-07-08 | 0.5103 | 0.7327 | 0.2198 | 0.9819 | 0.9440 | 0.9792 | 0.3750 | 0.2516 | 0.6050 | 0.4184 | 0.5653 | 0.8002 | 0.4581 |
| XltTs2Cgpp | 2020-07-08 | nan | 0.5207 | 0.5074 | 0.1620 | 0.1717 | 0.4180 | nan | nan | nan | nan | nan | nan | 0.7693 |
| hETU2yttQA | 2020-07-08 | 0.1306 | 0.0454 | 0.2603 | 0.8162 | 0.1992 | 0.3770 | 0.6795 | 0.3562 | 0.5801 | nan | nan | nan | 0.5390 |
| LGMy2anhFH | 2020-07-08 | 0.0273 | 0.1954 | 0.9173 | 0.0818 | 0.3180 | 0.2659 | 0.3705 | 0.3662 | 0.1917 | 0.1291 | 0.9909 | 0.4217 | 0.3193 |
| q0wQCNiPW0 | 2020-07-08 | 0.0155 | 0.7364 | 0.1586 | 0.1838 | 0.0673 | 0.9084 | nan | nan | nan | 0.0825 | nan | 0.1329 | 0.0851 |
| BdJzJyT1oh | 2020-07-08 | 1.0000 | 0.9739 | 0.9990 | 0.0776 | 0.0450 | 0.7628 | 0.2938 | 0.2972 | 0.2837 | 0.4417 | 1.0000 | 0.9986 | 0.8565 |
| dEL4EI3gFn | 2020-07-08 | 0.0888 | 0.2125 | 0.1539 | 0.4660 | 0.2859 | 0.8259 | 0.1356 | 0.2868 | 0.0685 | 0.2818 | 0.8167 | 0.3841 | 0.6334 |
| Y0zvIBNAtD | 2020-07-08 | 0.2220 | 0.2292 | 0.7280 | 0.0076 | 0.2678 | 0.6056 | 0.6352 | 0.5753 | 0.4457 | 0.4648 | 1.0000 | 0.8628 | 0.9358 |
| BljEJA8urJ | 2020-07-08 | 0.4565 | 0.1280 | 0.0874 | 0.2570 | 0.3195 | 0.5767 | nan | nan | nan | 0.8377 | 0.4537 | 0.5824 | 0.9497 |

Figure 6: Anomaly detection dashboard with a visualization of the scores from a collection of anomaly detection models.

In Figure 6, we display an example of what the anomaly detection dashboard looks like on a given day. Each row corresponds to a participant, and each column corresponds to a different anomaly detection score. The participant's identifier and the last date their scores were successfully updated is displayed, along with the anomaly scores for each feature. The score from a multivariate model is also displayed, which may capture interdependencies between features that affect whether or not a measurement is anomalous. For example, major disruptions in sleep naturally coincide with periods of long-distance travel (abnormally large radius of gyration). Note that cells are colored according to the severity of the scores. While this dashboard is mainly used for research at this point, if reliable anomalies are detected they could be promoted to be used on the clinician dashboard.

## 5 Example Analysis: Measuring the Effect of Singapore's "Circuit-breaker"

Due to the SARS-CoV-2 (COVID-19) pandemic, Singapore has imposed a stay-at-home order or *cordon sanitaire* which is formally called "the 2020 Singapore Circuit Breaker measures" or CB. This lockdown was in effect from 7 April 2020 until 1 June 2020, after which gradual stages of reopening have occurred. During this period, people were required to stay at home as much as possible, avoid non-essential travel and social visits, and to maintain social distancing in public. As a result of the lockdown, we would expect to see effects in some digital phenotyping features. As a test for our digital phenotyping system, we performed and report here a data comparison using 20 participants' data before and after the start of this "circuit breaker".

| Feature name | 6-wk mean before CB starts | 6-wk mean after CB starts | Paired t-test (p-value) | Wilcoxon signed rank test (p-value) |
|---|---|---|---|---|
| *SmartPhone features* | | | | |
| accel_L_std (L: length of the accel. vector) | 0.537945 | 0.386095 | 0.00019 | 0.00039 |
| accel_ddt_max (ddt: time derivative) | 0.008181 | 0.005644 | 0.00419 | 0.00102 |
| ambientLight.hourly_max_log1p_lux | 2.6379 | 2.2704 | 0.00055 | 0.00116 |
| callLog_Incoming Call | 0.65024 | 0.26900 | 0.00020 | 0.00029 |
| gps-mobility_Hometime / mins | 1070.3 | 1328.1 | 0.00037 | 0.00009 |
| gps-mobility_SigLocsVisited | 1.3939 | 1.1764 | 0.00002 | 0.00013 |
| powerState.hourly_n_screen_on | 6.0737 | 4.6965 | 0.01202 | 0.01237 |
| tapsLog.daily_n_unique_apps | 15.619 | 13.290 | 0.04055 | 0.04380 |
| tapsLog.daily_n_taps_in_entertainment | 270.68 | 357.99 | 0.06134 | 0.04004 |
| *Wrist-wearable features* | | | | |
| steps.daily_n_steps | 2527.4 | 1641.9 | 0.00600 | 0.00719 |
| steps.daily_n_mins_walk | 72.340 | 52.964 | 0.00409 | 0.00511 |
| heart.daily_HR_mean | 82.702 | 79.670 | 0.03473 | 0.01374 |
| heart.daily_HR_min | 55.653 | 54.549 | 0.14811 | 0.10843 |
| sleep_total_hrs | 8.927 | 8.469 | 0.29790 | 0.07932 |
| sleep_mean_efficiency | 93.296 | 92.483 | 0.03461 | 0.01237 |

Table 2: Comparison of 6-week digital phenotyping data before (from 45 days before to 3 days before) and after (from 3 days after to 45 days after) Singapore's Circuit-Breaker (CB) was instituted on 7 April 2020. Data recorded for participants with complete data both before and after the start of circuit breaker.

Table 2 shows a subset of features that show statistically significant difference before and after the circuit breaker was instituted on 7 Apr 2020. Not surprisingly, since people were required to stay at home, the home-time has increased and the number of significant locations visited has decreased. Features related to physical activity (heart-rate, steps, and acceleration) have also decreased as might be expected. Both sleep and sleep efficiency have also decreased amongst these participants. It is also noteworthy that participants appear to use a fewer number of apps, perhaps because there is no need for some apps such as maps for navigation or those checking bus arrival times; however, it appears they spend more time in entertainment apps. Moreover, the ambient light indoors is generally dimmer than it is outdoors, and so the observed decrease in maximum ambient light is also as expected.

We compared our results with another study based on Fitbit use, the Health Insights Singapore (hiSG) study [37]. Daily steps count decreased by ~35% in our study and ~42% in hiSG; the minimum heart rate decreased by 1.1 bpm in our study and the resting heart rate decreased by 1.6 bpm in the hiSG study; sleep efficiency decreased by 0.8% in our study and by 0.2% in the hiSG study. All comparisons between both studies were consistent in demonstrating changes before and after onset of circuit breaker measures in Singapore.

# 6   Conclusion

Digital phenotyping is a promising area in healthcare but requires great care and effort in designing a system that is easy to use, safe in terms of data security and privacy, and collects data with enough details and reliability to be useful in research and patient care. We found the *Beiwe* platform to be a suitable base that we could use and extend to create the HOPES platform. Our main extensions have been adding many more data sources for collection and in integrating the use of a wearable, and the development of a large set of monitoring and participant management platforms.

We were also driven by meeting all the requirements of a clinical research study for schizophrenia (HOPE-S). This required us to develop significant enhancements in security, privacy, ease-of-use and scalability, choosing a careful combination of public cloud and on-premises operation.

Since massive amounts of diverse data are collected and in digital phenotyping, we have had to create new mechanisms to clean, process, present, explore and analyze data. These need to serve the needs of clinical research study operations, clinical care, platform developers and researchers, hence a range of platforms and data platforms have been developed.

Our initial platform is in use in a clinical trial (HOPE-S) and interim results will soon be reported. An initial test using SARS-CoV-2 as a test-case yielded meaningful and expected results consistent with expected lockdown behaviors, and was consistent with an independently conducted study in the same country.

**Acknowledgements**

We thank our colleagues at Singapore's Institute for Mental Health, who have been involved in defining and testing the HOPES system, and leading the HOPE-S clinical study. We also thank Fitbit for providing technical assistance on using their cloud API.

# References

1. Marsch LA. Digital health data-driven approaches to understand human behavior [published online ahead of print, 2020 Jul 12]. *Neuropsychopharmacology*. 2020;1-6. doi:10.1038/s41386-020-0761-5
2. McGinnis JM, Williams-Russo P, Knickman JR. The case for more active policy attention to health promotion. *Health Aff (Millwood)*. 2002;21(2):78-93. doi:10.1377/hlthaff.21.2.78
3. Schroeder SA. Shattuck Lecture. We can do better--improving the health of the American people. *N Engl J Med*. 2007;357(12):1221-1228. doi:10.1056/NEJMsa073350
4. Torous J, Kiang MV, Lorme J, Onnela JP. New Tools for New Research in Psychiatry: A Scalable and Customizable Platform to Empower Data Driven Smartphone Research. *JMIR Ment Health*. 2016;3(2):e16. Published 2016 May 5. doi:10.2196/mental.5165
5. National Steps Challenge™. https://www.healthhub.sg/programmes/37/nsc. Accessed August 19, 2020.
6. Waddell K, Shah P, Adusumalli S, Patel M. Using Behavioral Economics and Technology to Improve Outcomes in Cardio-Oncology. *JACC: CardioOncology*. 2020;2(1):84-96. doi:10.1016/j.jaccao.2020.02.006
7. Cote DJ, Barnett I, Onnela JP, Smith TR. Digital Phenotyping in Patients with Spine Disease: A Novel Approach to Quantifying Mobility and Quality of Life. *World Neurosurg*. 2019;126:e241-e249. doi:10.1016/j.wneu.2019.01.297
8. Wright AA, Raman N, Staples P, et al. The HOPE Pilot Study: Harnessing Patient-Reported Outcomes and Biometric Data to Enhance Cancer Care. *JCO Clin Cancer Inform*. 2018;2:1-12. doi:10.1200/CCI.17.00149
9. Saeb S, Zhang M, Karr CJ, et al. Mobile Phone Sensor Correlates of Depressive Symptom Severity in Daily-Life Behavior: An Exploratory Study. *J Med Internet Res*. 2015;17(7):e175. Published 2015 Jul 15. doi:10.2196/jmir.4273
10. Meyer N, Kerz M, Folarin A, et al. Capturing Rest-Activity Profiles in Schizophrenia Using Wearable and Mobile Technologies: Development, Implementation, Feasibility, and Acceptability of a Remote Monitoring Platform. *JMIR Mhealth Uhealth*. 2018;6(10):e188. Published 2018 Oct 30. doi:10.2196/mhealth.8292
11. Barnett I, Torous J, Staples P, Sandoval L, Keshavan M, Onnela JP. Relapse prediction in schizophrenia through digital phenotyping: a pilot study. *Neuropsychopharmacology*. 2018;43(8):1660-1666. doi:10.1038/s41386-018-0030-z
12. Mindstrong Health. https://mindstronghealth.com/. Accessed August 19, 2020.
13. Lief Therapeutics. https://getlief.com/. Accessed August 19, 2020.
14. QuantActions. https://quantactions.com/. Accessed August 19, 2020.
15. Jaimini U, Thirunarayan K, Kalra M, Venkataraman R, Kadariya D, Sheth A. "How Is My Child's Asthma?" Digital Phenotype and Actionable Insights for Pediatric Asthma. *JMIR Pediatr Parent*. 2018;1(2):e11988. doi:10.2196/11988
16. Faherty LJ, Hantsoo L, Appleby D, Sammel MD, Bennett IM, Wiebe DJ. Movement patterns in women at risk for perinatal depression: use of a mood-monitoring mobile application in pregnancy. *J Am Med Inform Assoc*. 2017;24(4):746-753. doi:10.1093/jamia/ocx005
17. Spinazze P, Bottle A, Car J. Digital Health Sensing for Personalized Dermatology. *Sensors (Basel)*. 2019;19(15):3426. Published 2019 Aug 5. doi:10.3390/s19153426

18. UCSF TemPredict Study. Oura Ring. https://ouraring.com/ucsf-tempredict-study. Accessed August 19, 2020.
19. Hemphill NM, Kuan MTY, Harris KC. Reduced Physical Activity During COVID-19 Pandemic in Children With Congenital Heart Disease. *Can J Cardiol*. 2020;36(7):1130-1134. doi:10.1016/j.cjca.2020.04.038
20. Lee XK, Chee NIYN, Ong JL, et al. Validation of a Consumer Sleep Wearable Device With Actigraphy and Polysomnography in Adolescents Across Sleep Opportunity Manipulations. *J Clin Sleep Med*. 2019;15(9):1337-1346. doi:10.5664/jcsm.7932
21. Health Outcomes Via Positive Engagement in Schizophrenia - Full Text View - ClinicalTrials.gov. Clinicaltrials.gov. https://clinicaltrials.gov/ct2/show/NCT04230590. Accessed August 19, 2020.
22. Beiwe Research Platform. https://www.beiwe.org/. Accessed August 19, 2020.
23. onnela-lab. GitHub. https://github.com/onnela-lab. Accessed August 19, 2020.
24. Schueller SM, Begale M, Penedo FJ, Mohr DC. Purple: a modular system for developing and deploying behavioral intervention technologies. *J Med Internet Res*. 2014;16(7):e181. Published 2014 Jul 30. doi:10.2196/jmir.3376
25. Purple Robot - CBITs TECH Web Site. https://tech.cbits.northwestern.edu/purple-robot/. Accessed August 19, 2020.
26. cbitstech/Purple-Robot. GitHub. https://github.com/cbitstech/Purple-Robot. Accessed August 19, 2020.
27. AWARE – Open-source Context Instrumentation Framework For Everyone. https://awareframework.com. Accessed August 19, 2020.
28. AWARE Framework. GitHub. https://github.com/awareframework. Accessed August 19, 2020.
29. Matcham F, Barattieri di San Pietro C, Bulgari V, et al. Remote assessment of disease and relapse in major depressive disorder (RADAR-MDD): a multi-centre prospective cohort study protocol. *BMC Psychiatry*. 2019;19(1):72. Published 2019 Feb 18. doi:10.1186/s12888-019-2049-z
30. RADAR-base. GitHub. https://github.com/RADAR-base. Accessed August 19, 2020.
31. RADAR-CNS. https://www.radar-cns.org. Accessed August 19, 2020.
32. Dagum P. Digital biomarkers of cognitive function. *NPJ Digit Med*. 2018;1:10. Published 2018 Mar 28. doi:10.1038/s41746-018-0018-4
33. Insel TR. Digital Phenotyping: Technology for a New Science of Behavior. *JAMA*. 2017;318(13):1215-1216. doi:10.1001/jama.2017.11295
34. LeGates TA, Fernandez DC, Hattar S. Light as a central modulator of circadian rhythms, sleep and affect. *Nat Rev Neurosci*. 2014;15(7):443-454. doi:10.1038/nrn3743
35. Box G E P., Jenkins G M., Reinsel G C., Ljung G M. Time Series Analysis: Forecasting and Control. 5th ed. Hoboken, New Jersey: John Wiley & Sons; 2015.
36. Rasmussen C, Williams C. *Gaussian Processes For Machine Learning*. Cambridge, Mass: MIT Press; 2008.
37. Ong JL, Lau TY, Massar SAA, et al. COVID-19 Related Mobility Reduction: Heterogenous Effects on Sleep and Physical Activity Rhythms, ArXiv Preprint: 2006.02100v2

# Supplementary Material: HOPES - An Integrative Digital Phenotyping Platform for Data Collection, Monitoring and Machine Learning

Xuancong Wang[1], Nikola Vouk[1], Creighton Heaukulani[1], Thisum Buddhika[1], Wijaya Martanto[1], Jimmy Lee[2,4], Robert JT Morris[1,3]

[1]Ministry of Health Office for Healthcare Transformation (MOHT), Singapore
[2]Institute of Mental Health, Singapore
[3]National University of Singapore, Singapore
[4]Lee Kong Chian School of Medicine, Nanyang Technological University, Singapore

Corresponding Author:  Creighton Heaukulani (creighton.heaukulani@moht.com.sg)

Abstract

This paper provides supplementary material for our main paper (titled "HOPES - An Integrative Digital Phenotyping Platform for Data Collection, Monitoring and Machine Learning") in which we describe the development of, and early experiences with, a comprehensive Digital Phenotyping platform: **H**ealth **O**utcomes through **P**ositive **E**ngagement and **S**elf-Empowerment (HOPES). HOPES is based on the open-source *Beiwe* platform but adds a much wider range of data collection, including the integration of wearable devices and further sensor collection from the smartphone. Requirements were in part derived from a concurrent clinical trial for schizophrenia. This trial required development of significant capabilities in HOPES for security, privacy, ease-of-use and scalability, based on a careful combination of public cloud and on-premises operation. We describe new data pipelines to clean, process, present and analyze data. This includes a set of dashboards customized to the needs of research study operations, and for clinical care. A test use case for HOPES is described by analyzing the digital behavior of 20 participants during the SARS-CoV-2 pandemic.

## 1. HOPES Solution Architecture

The HOPES platform infrastructure was separated out into functions of administration, data upload, data encryption, wearable data collection, operational management. Figure 1 shows the separation of the layers between the cloud environment and the on-premise data analytics environment. The data collection, monitoring, and aggregation infrastructure are separated into logical networks that share common encrypted storage, and access to appropriate encryption keys to ensure the data collected is consistent. Access to the encrypted study data, and the study decryption keys is exclusively provided to authorized data analytical services on the R&D Premise for processing. The collection of wearable data from cloud data collection sources is independent of the collection of HOPES application data, but both are normalized into a common set of formats and encrypted using the public key of each respective participant's data to ensure consistent data processing. The private key is only accessible to authorized data analytics pipelines.

**HOPES SOLUTION ARCHITECTURE**

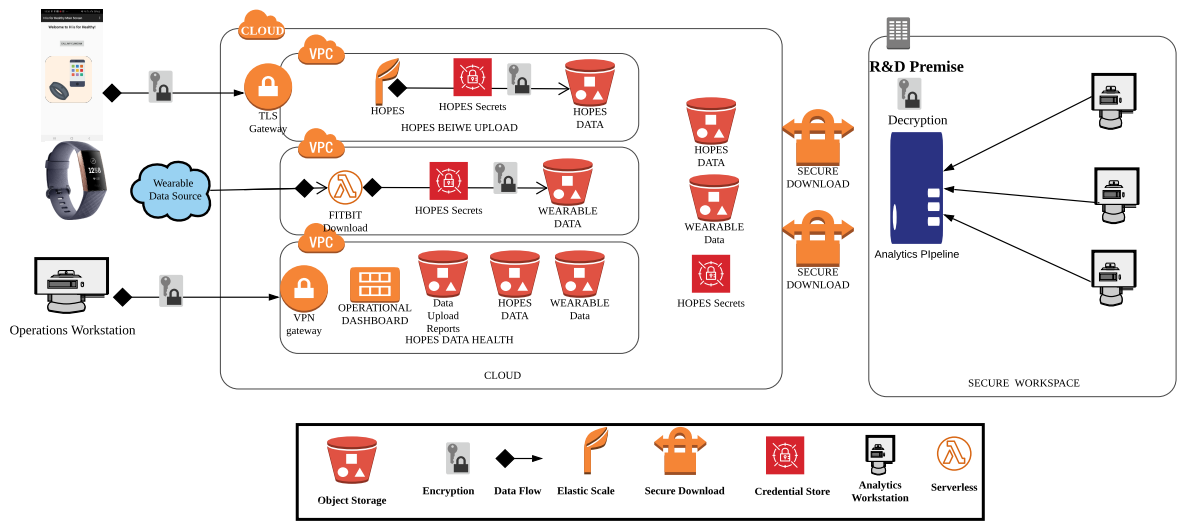

Figure 1: HOPES Multi-Tier Scalable Distributed Architecture

The solution has separated administration to only authorized administrators accessing through private VPN connections. Monitoring dashboards are separated into a different logical network accessible through a different VPN gateway, and only metadata is visible to operators. Secure download is made possible through private authorized connections, and secured credentials. Once data is processed by the analytics pipeline, de-identified processed data is made available to data scientists and clinicians for further analysis visible in the data exploration tooling, anomaly dashboard, clinician dashboard, and as raw information for further analysis. Data is encrypted at all times while in the data collection infrastructure, and only decrypted during the analytics pipeline.

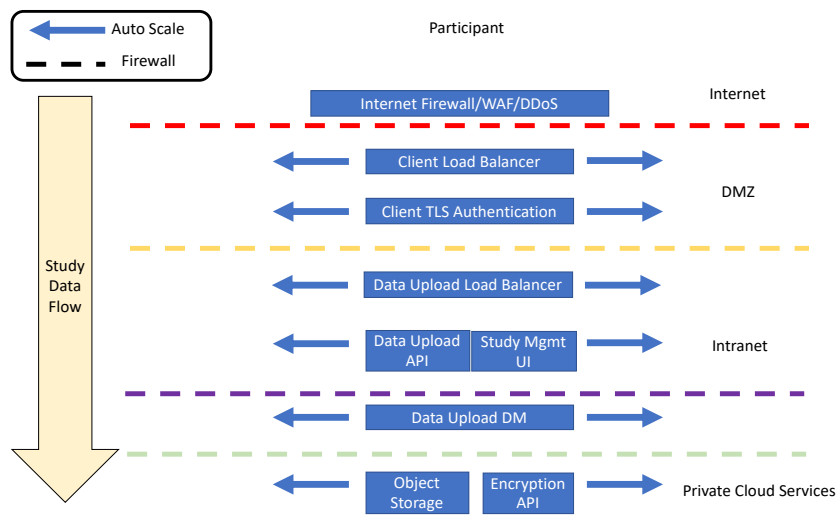

Figure 2 Scalable Infrastructure

Figure 2 represents the data collection infrastructure enhanced to leverage the best cloud and security architecture through separation of capability from data collection, data administration, as well as auto-scale and load balancing at every level. The infrastructure is stateless transaction-to-transaction. The architecture implements a secure certificate-based authentication and a rotating credential to ensure only authorized valid participants are connecting into the infrastructure. Automated patching, web application firewalls, distributed-denial of service protection, credential vaults, and secure software development practices ensure a robust integrity for the infrastructure. An automation framework was written to deploy the software reliably and securely.

## 2. Additional App Enhancement
### 2.1 QR Scanning for Participant Registration

To facilitate user registration, we implemented a QR code reader to replace manual entry of credentials. To register a participant's phone to the study server, the registration information is generated on the server side with an asymmetric encryption and decryption key. The registration information together with the encryption key is stored in the QR code and can be sent to the participant via email or as a print-out. A migration feature was added to allow for users to migrate to new study phones and maintain the de-identification credentials as well as maintain integrity of study data.

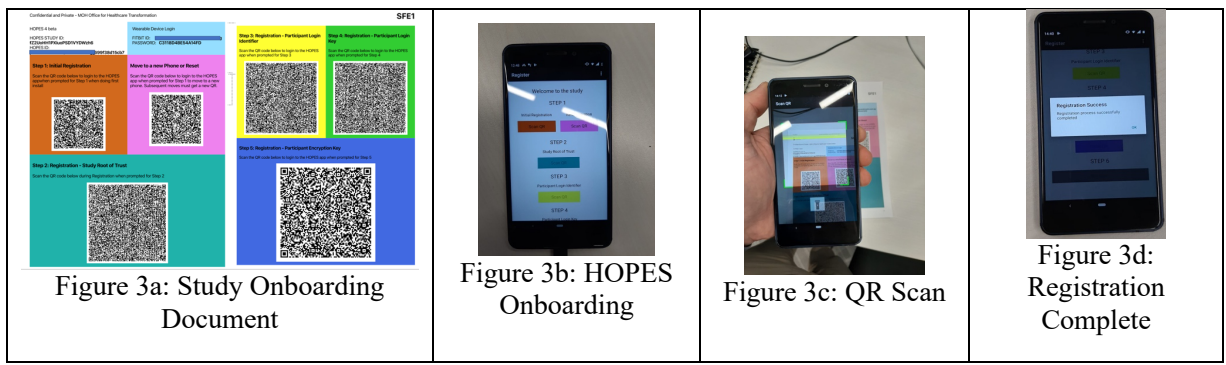

Figure 3a: Study Onboarding Document | Figure 3b: HOPES Onboarding | Figure 3c: QR Scan | Figure 3d: Registration Complete

As shown in Figure 3(a-d), the participant's onboarding sheet is securely generated by the study administrators, passed to the clinicians for onboarding, and then when participants are ready to be onboarded, the sheet is used to configure the HOPES app by scanning the QR codes. The onboarding consists of six color-coded steps used to input the necessary information for logging into the server, defining the data randomization, and connecting into the infrastructure. For simplicity, and the application ensures each step is followed in order. The QR codes contain the randomized de-identification information, the encryption keys for encoding the data, as well as secret key to connect into the infrastructure. In the latest version, the QR codes are encrypted to ensure confidentiality of the information offline. The keys are only valid during the study duration, and are invalidated upon participant completion of the study, or the study completes. This information is never made available during processing, and securely stored offline by the study administrators, and destroyed after the study is complete.

This process has allowed us to onboard participants easily without the error of inputting credentials, server addresses, facilitate self-service onboarding, and simple onboarding for our participants.

## 2.2   HOPES Debug Interface

The debug console can help test almost every functionality of the app directly on the phone without the need to connect to a computer running Android Studio. It logs every feature that can be collected. It serves two main purposes: for technical troubleshooting (certain features may work differently on certain brands of phones with certain Android versions) and to give more technically-oriented users or inquirers some degree of privacy assurance by showing what exact information is collected and sent to the server. For example, Figure 4 shows an example of the taps log being captured. The debug console can connect to every feature logger. Data from that feature's listener can be displayed in the console before being encrypted and written to file. The debug console is directly accessible when no study has been registered. After study registration, it will be locked by a password or totally disabled depending on the study settings. This is to prevent interference to the data collection for the study.

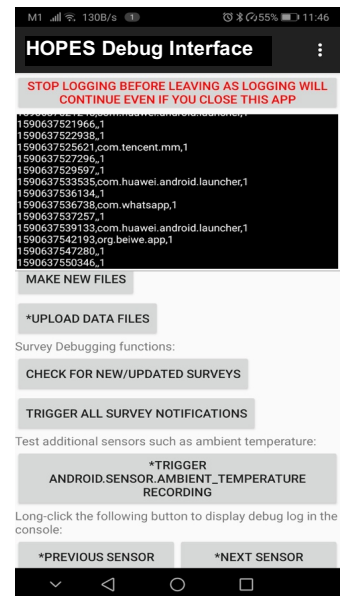

Figure 4: the HOPES Debug Interface

## 3.   Newly-Added Digital Phenotyping Features

Below, we describe features that are newly added or enhanced on top of the Beiwe distribution, expanding on those already stated in the main paper.

i.   Sensor: Pedometer

Although step counts are readily captured by most wrist-wearable devices like Fitbit, it is still beneficial to capture step counts on the phone, since some people do not wear smart bands or watches. For those that do wear wrist devices, the differences between the two sources of step data can provide some interesting information. For example, if the wrist device registers steps during a particular period but phone does not, this may suggest that the user is merely walking around their home or office (i.e., not travelling), in which case their phone may have been left on a desk or table.

ii.   Sensor: Ambient Light

We capture ambient light since studies have suggested there is correlation between a patient's mental health and their preferred environmental lighting. Additionally, the ambient light in a person's sleep environment would likely affect sleep quality, which may in turn have an influence on their mental wellness. In our implementation of the "study settings", the researcher can set a time interval during which the ambient light reading is taken.

iii.   Sensor: Magnetometer

The magnetometer returns the direction of magnetic fields passing through the phone. This information tells us about the orientation and alignment of the phone, which in turn can determine the rotational motion of the phone.

iv.   Capturing Taps

Taps provide two types of information that may be related to a person's health. The speed at which a person taps may give a hint of their wellbeing; for example, a fatigued person may tap more slowly, or some diseases may cause small, uncontrollable movements. The apps a person uses (determined from their taps) also gives an indication of their behavior. For example, a relapsing schizophrenia patient may have significantly altered communication, reflected in the number and speed of taps he/she made in each app. We make use of Android application overlay to capture taps. In particular, it creates an invisible tiny popup window on the screen and watches for every tap outside the window. It then queries the Android Usage Stat Manager to get the app name in which the last tap was made. We record the timestamp, app-name, and screen orientation of the phone for each tap.

    v.    Accessibility Taps

The taps-capturing method (described in Section 4.1.4) using the Android application overlay cannot capture Android system buttons such as Home, Back, and Recent Apps. Additionally, it cannot capture more detailed information about the button that has been tapped due to Android's privacy preserving strategies. We attempt to capture typing error rates, which we believe can be affected by a person's the physical or mental condition. We can determine this from how often the DELETE key on the keyboard is tapped. To measure tapping speed, we also need to know whether the person is typing on the keyboard or navigating in a social messaging app.

    vi.    Sociability Messages

For our study, changes in a participant's sociability (i.e., their communication with others) may be related to his/her mental health status. This may be reflected in their activity in social messaging apps. In particular, the number of incoming and outgoing messages, the number of message senders to which the participant has replied, and the lengths of outgoing messages, for example, may indicate a degree of social engagement. The original Beiwe app captures incoming and outgoing SMS messages. However, in Singapore most people use social messaging apps like WhatsApp as their primary method for text communication. We therefore make use of the Android Accessibility service privilege to acquire message meta data from social messaging apps. We have so far only implemented this for WhatsApp, but it can be easily extended to other social messaging apps in the future. To protect user privacy, we do not capture the content of message, but only the meta data of each message, i.e., the timestamp, direction (incoming or outgoing), sender/receiver hashed identity (not their names), message length, and message type (image, text, voice, etc.).

    vii.    Sociability Calls

Similar to Sociability Messages, our app also leverages the Android Accessibility service privilege to capture calls within social messaging apps since, in Singapore, a significant number of phone calls are made using social messaging apps rather than the phone's SIM card. Following the data format of the call log feature implemented in Beiwe, the HOPES app records the timestamp, duration, direction (incoming or outgoing), type (voice or video call), and hashed sender/receiver identity of every call within WhatsApp. Depending on the need, this can be easily extended to other social messaging apps.

    viii.    Time Zones

In general, changes in time zone can have a significant impact on both one's physical and mental health, as well as one's usage data. If not handled properly, it can cause large discrepancies in clinical predictions for participants who often travel across time zones. For every feature that we capture, we have therefore added one additional field recording the current system time zone.

# 4. Additional Information on HOPES Fitbit Component

## 4.1 Fitbit Setup

We used the Fitbit Charge 3 in our study. User accounts were pre-created with system generated email addresses and passwords. Separate "main" application was created in order to download data from user accounts, following "Authorization Code Grant Flow". Data was downloaded using the Fitbit Web API, in terms of intraday time series.

## 4.2 Fitbit Data Download Architecture

We used *AWS lambda*, to extract user data from the Fitbit cloud, update access tokens, write encrypted data to files, calculate meta data, and then upload into *AWS S3 buckets*. The Fitbit data downloader lambda is triggered every hour by a rule setup in *AWS CloudWatch.* Access token of each user was setup in the *AWS Secret Manager* and updated as necessary.

# 5. Back-end Data Processing Pipeline

The data processing back-end is designed to reformat and process data for downstream machine learning models. We created a master script called *./periodic-run.sh* which is scheduled to run periodically with a configurable time interval. Sequentially, it will:

1. Decrypt new data files (output to *./decrypted*). In the original system design, for every participant, for every feature, one file will be generated every hour. The file is uploaded and stored on server in encrypted form. The backend data processing server keeps all of these files in both encrypted and decrypted format (for decryption troubleshooting). The script will recursively decrypt every file in the directory if their decrypted version does not exist.

2. Patch new data files (output to *./1.decrypted*). This will fix some file format inconsistency issues, e.g., the timestamp in some device feature files are in units of seconds rather than the milliseconds used in phone features.

3. Concatenate all files (output to *./2.decrypted*) for the same participant and feature category, fix corrupted files, ignore duplicate files, etc.

4. Fix some data value issues (output to *./3.decrypted*), such as clamping the ambient light value range (on certain phones, the ambient light sensor return lux values like 1e+15 upon saturation, while the authentic maximum value is around 65k), fix duplicate and corrupted data entries, and add the app-group column, see Section 4.3.2 for details). As compared to Step 2, this steps fixes issues after file concatenation.

5. Summarize into high-level features (output to *./4.decrypted*). The raw data comes as irregularly-spaced time-series data. We use conventional methodology to extract high-level regular time-series features from this raw data. We typically extract summary statistics such as the maximum, minimum, mean, median, standard deviation, count, etc., of some features over at an hourly or daily sampling interval. For the detailed specification, see Table 1 and Section 5.1.

6. Concatenate all features and combine all data files (output to *./5.decrypted*). Some of the features are hourly features and some are daily features. For hourly features, we concatenate the vectors from all hours in one day to form a feature for that day. Next, the system first combines every participant's daily feature data into one compressed file, and it then combines all participants'' data into a single (rather large), compressed file.

| Feature Name | Source | # of dims. daily | # of dims. hourly | Total Dimensions |
|---|---|---|---|---|
| Sleep | Fitbit watch | 17 | 0 | 17 |
| Steps | Fitbit watch | 14 | 4 | 110 |
| Heart rate | Fitbit watch | 12 | 6 | 156 |
| GPS mobility | smartphone | 15 | 0 | 15 |
| Accelerometer | smartphone | 0 | 5 | 120 |
| Accessibility taps | smartphone | 0 | 1 | 24 |
| Call logs | smartphone | 5 | 0 | 5 |
| Ambient Light | smartphone | 3 | 2 | 51 |
| Power state | smartphone | 6 | 2 | 54 |
| Sociability call log | smartphone | 4 | 0 | 4 |
| Sociability msg log | smartphone | 9 | 0 | 9 |
| Taps | smartphone | 13 | 6 | 157 |
| SMS | smartphone | 7 | 0 | 7 |
| Total | (both) | 105 | 26 | 729 |

Table 1: Summary of all the extracted and processed high-level regular time-series features (dims, or dimensions) from raw digital phenotyping features

Our design of the data processing pipeline caters to contemporary machine learning (ML) models. Depending on the model and the scale of the study, any stage of output from *./3.decrypted* to *./5.decrypted* can be directly plugged into any downstream analysis. For example, in Step 5, there will be information loss due to feature summarization, so some ML models may preferably use a 2-stage or multi-stage recurrent neural network (RNN), in which the first-stage RNN reads the raw event time-series data, one RNN for each kind of feature, followed by subsequent RNNs combining the output of all those first-stage RNNs to predict an output. Such a model would take its inputs from *./3.decrypted*.

### 5.1 Specification for High-level Feature Extraction

The purpose of this specification is to convert the raw data into regular daily time-series data, i.e., one fixed-dimensional vector every day for every participant, and to ensure value continuity (e.g., sleeping at 23:59 and 00:01 are very close, so should be reflected in the values as well) for downstream machine learning models.

1. Accelerometer (accel.csv)
   INPUT: every 10 minutes (or longer), we have 0-3500 triplets of (x, y, z)
   OUTPUT: every hour, max/min/std/mean of $\sqrt{x^2 + y^2 + z^2}$; and max rate of change of acceleration, i.e., $\max\left(\begin{vmatrix}\Delta x/\Delta t\\ \Delta y/\Delta t\\ \Delta z/\Delta t\end{vmatrix}\right)$.

2. Accessibility (accessibilityLog.csv)
   OUTPUT: every hour, the total number of taps; the number of keyboard taps; the number of DELETE key taps; the ratio of (# of DELETE key taps)/(# of keyboard taps)

3. Call log (callLog.csv)
   INPUT: a log of every SIM phone call, with timestamp, incoming/outgoing, hashed identity and call duration
   OUTPUT: [same as sociabilityCallLog, wherever applicable]

4. Location (locTime-gps.csv)
   [We follow Beiwe-Analysis sample code.]

5. Heart rate (heart.csv)
   INPUT: every 5 seconds, the heart-rate (HR) value in beats per minute, (if not wearing, take the previous value)
   OUTPUT: every hour, max/min/std/mean of HR, max/min/std of HRV, (i.e., $\Delta HR/\Delta t$); every day, max/min/std/mean/median of HR, max/min of HRV, mean/std of abs(HRV), 25% and 12.5% quantile of HR (since minimum heart-rate can be quite volatile).

6. Ambient light (light.csv)
   INPUT: every 5 minutes (or longer), a value in units of lux ranging from 0 to ~50k
   OUTPUT: every hour, max/mean/min of $\log(1 + value)$, 50-high value (the mean of the top 50% of values)

7. Power state (powerState.csv)
   INPUT: a log of every event such as screen turn on/off, power-down signal, device idle state change, etc.
   OUTPUT: every hour, the number of seconds while the screen is on and the total number of power events (screen-on, screen-off and power-off); every day, the number of power-down signals, max/min/std/mean duration (in seconds) of each screen-on session

8. Sleep (sleep.csv)
   INPUT: every day, during estimated sleep hours, a time log of the type (deep, light, REM and awake) and duration of every sleep segment (with 30s resolution)
   OUTPUT: every day (from 15:15 to 15:15 the next day), the total duration of deep/light/rem/awake segments respectively; the number of awake segments inside the main sleep; the number of awake segments ≥ 180 seconds inside the main sleep; the ratio of (total duration of deep/light/REM/awake segments inside main sleep)/(total duration of main sleep) respectively; the starting and ending time of the main sleep with respect to 7:15 and 23:15; time-to-asleep (time from night going to bed to falling asleep) and time-to-getup (time from waking up in the morning to getting out of bed); sleep efficiency.

9. Sociability call log (sociabilityCallLog.csv)
   INPUT: a log of every WhatsApp call event
   OUTPUT: every day, number of incoming calls; number of outgoing calls; number of missed calls; total duration of calls; number of people talked (either incoming or outgoing, but must have duration $> 0$)

10. Sociability message log (sociabilityLog.csv)
    INPUT: a log of every WhatsApp message event

OUTPUT: every day, number of received messages; number of sent messages; total length of received messages; total length of sent messages; number of contacts who has 1) only received message, 2) both received and sent messages, and 3) only sent messages, respectively.

11. Steps (steps.csv)
    INPUT: every minute, number of steps made
    OUTPUT: every hour, total number of steps, max number of steps in a minute within that hour; every day, total number of steps, # of wearing minutes, # of minutes with steps, # of walks (a walk is at least 3 consecutive minutes with each minute having at least 10 steps), the max/mean steps during each walk, the max/mean duration of each walk, the average steps per minute during each walk, max number of consecutive minutes each with more than 3 steps and 30 steps (to predict whether have gone outdoor).

12. Taps (tapsLog.csv)
    INPUT: a log of every tap, with timestamp, screen orientation and in-app name
    OUTPUT: every hour, the total number of taps, the max/min/std/mean/median of inter-tap duration while screen is on; every day, the number of distinct in-app names, the number of taps made in each APP category (games, social media, social messenger, etc.), and the max/min/std/mean/median of inter-tap duration in social messenger.

13. SMS log (textsLog.csv)
    INPUT: a log of every SMS, with timestamp, send/receive, hashed identity and message length
    OUPUT: [same as sociability message log, wherever applicable)

## 5.2 Android App Grouper

A person's behavior and mental health state is to some extent reflected in the kind of apps they spend their time in. We developed an app-grouper which classifies all apps into 7 groups, according to each app's package name and Google Play Store's categorization of the app.
1. **Social Messenger**: apps that involve person-to-person real-time communication, e.g., *WhatsApp, QQ, Wechat, Telegram, Dating apps, Email*, etc.
2. **Social Media**: apps that involve information exchange with the world, e.g., *Youtube, Chrome, Firefox, Facebook, Instagram, Twitter*, etc.
3. **Entertainment**: all apps related to education, comics, stories, sports, music and video, etc., for personal life entertainment but are not classified under games
4. **Map Navigation**: apps that are related to travel, map and navigation, e.g., *SGBus, SG BusLeh, Maps, MAPS.ME*, etc.
5. **Utility Tools**: apps that are used for various utility purposes such as banking, finance, document/photo/video viewer and editors, e.g., *PayLah, PayNow, OCBC, AXS, Dropbox, WPS Office, PDF reader*, etc.
6. **Games**: all apps in the game category of the Play Store
7. **Android System**: all built-in apps that cannot be found in the Play Store, e.g., Phone call, SMS, Camera, Gallery, Contacts, Settings, etc. Take note that for some of the apps, their package names might not be the same across all brands of phones, thus, we decide not to further classify them.

# 6. Data Visualization Toolkit

The data visualization toolkit has been developed as a web interface using *Jupyter notebook*. The dashboard is highly configurable. The user needs to specify an input root path, inside which the directory structure can be either '*RootPath/StudyName/PatientName/FeatureName/timestamp-files*' or '*RootPath/StudyName/PatientName/FeatureName.csv(.gz)*', i.e., it can view data files both before and after concatenation (as described in Section 4.3). In the *master configuration dashboard* (as shown in Figure 5), the user can choose which study, participant, and feature to view. In other checkboxes, drop-down lists, and sliders, the user can set various options and choose different types of graphs to plot. For example, given the heart-rate data (which is in the form of a time-series with a heart-rate value every 5 seconds), the user can plot the maximum, minimum, mean, etc. for every interval of 1 week, 1 day, 1 hour, etc. You can also choose different columns in the CSV file. This can be plotted in the form of a line plot, a bar plot, or a scattered plot, among others. The example in Figure 5 shows a box-plot of heart-rate every day.

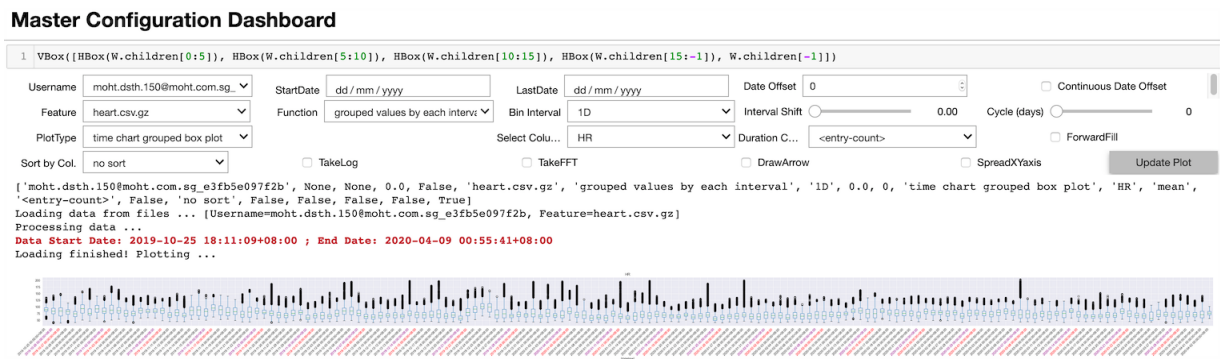

Figure 5: In the master configuration dashboard, you can select different data and choose various graphs and different plot options to plot.

The main advantage of this toolkit is that it is highly customizable. In practice, users often want to display a fixed set of specific graphs on some specific data. Instead of clicking through various control items in the *master configuration dashboard* every time, they only need to do that once, copy and paste the configuration parameters (in Figure 5 immediately below the *Update Plot* button) into their custom scripts when calling the *draw* function. An example is shown in the overview dashboard Figure 6. This toolkit is designed to be general-purpose. The user can also manually load individual CSV files, not necessarily from this study, and visualize them. The only requirement on the CSV files is that it must contain a column called *timestamp* or *datetime*.

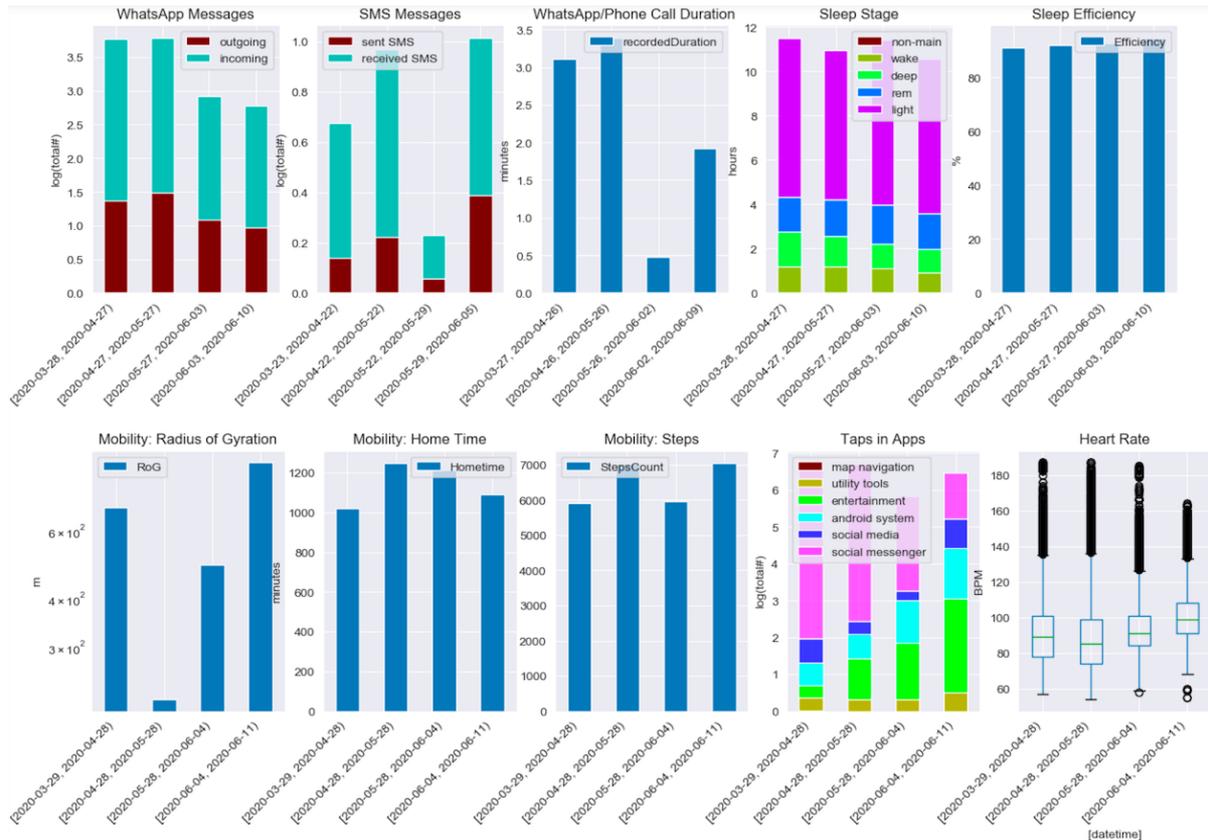

Figure 6: The overview dashboard, showing a particular patient's data in many feature aspects over a period of time.

## 7. Miscellaneous Platform Improvement

We have also modified the system so that a clinician's number can be set in the study settings. If it is not set in the study settings, then a '*PhoneNumberEntryActivity*' page will pop-up during the phone's registration. Since some studies may have participants that have already consented, they should not need to consent again in the app. In the study settings, therefore, one may leave the '*text that will appear in the app's consent form*' field empty, in which case the app will skip the consent page during the phone's registration. In the prototype backend, the administrator/researcher can also add a remark for every participant. The color of the '*Last Upload Time*' column can be customized in the study settings via a small Python script. The same applies to the '*Data Completion Status*' column. Moreover, clicking on a particular participant's '*Data Completion Status*' entry will produce an in-page pop-up window showing a table of the number of data records for that participant, for every feature, on every day. An example is shown in Figure 7.

Figure 7: Prototype back-end console showing the data completion details of the selected participant for every feature on every day

## 8. Dashboards

### 8.1 Data Collection Dashboard

This section gives a detailed explanation on what are the fields shown on the data collection dashboard.

From the phone:
1. *Location (last recorded hours ago)*: Time elapsed since the last location recorded until now
2. *Sociability (last recorded hours ago)*: Time elapsed since the latest WhatsApp message/WhatsApp call/SMS or a call
3. *Taps in Apps (last recorded hours ago)*: Time elapsed since the last touch on any application of the phone
4. *Last HOPES Uploaded*: Last time the system received data from the phone

From the Fitbit:
1. *Last Fitbit Uploaded*: Last time the system received data from the Fitbit
2. *Sleep (last uploaded hours ago)*: Time elapsed since the latest awake time after the last sleep

Other fields used in the dashboard:
1. *Participant*: Anonymous identifier given to the participant
2. *Last Clinic Visit (visit #)*: Last clinic visit day with the visit number

3. *Avg Fitbit Wearing Per Day (since last visit)*: Average Fitbit wearing time since the last visit, by using the heart rate which is captured in every 5 seconds. Time window for the calculation is from midnight of the clinic visit day to midnight of the day before the current day.
4. *Payment Progress (since last visit)*: Calculated based on the total Fitbit wearing time from the midnight of the last clinic visit day to the dashboard generation time. A day is considered as 22 hours long based on the requirement that a person should wear the Fitbit for at least 22 hours (i.e., this is considered 100% wearing-time). Payment progress refers to the participant's progress towards an inconvenience payment that may be provided to some participants in a clinical trial.
5. *Phone Model*: Model of the phone
6. *Enrollment*: Onboarding date of the participant

Color codes:

To highlight the *issues* in *data-health*, a color code is used to make it easy to spot.
- *Green*: Normal state. Latest data received within the last 24 hours
- *Orange*: Standby state. Latest data received within the last 96 hours
- *Red*: Alert state. No data received within the last 96 hours

Issue summary:

Data-health issues that are red and orange are shown based on the issue type which are *Phone Sync Issues, Fitbit Sync Issues, Sleep Data Issues*

Other fields:
- *Last Fitbit Data Downloaded*: The last time the system downloaded Fitbit data from the Fitbit cloud
- *Dashboard Generated*: Time of the current view of the dashboard generated

## 8.2 Clinician Dashboard

This section describes the features used in the clinician dashboard in detail

***Sleep***: The sleep graph is drawn based on the sleep duration (total time in bed) and sleep efficiency (ratio between sleep time and time in bed) directly retrieved from Fitbit. The vertical axis shows the sleep duration in hours while the horizontal axis indicates the duration for which the graph is drawn (see description of CW, PW, and PM below). The color of the graph depicts the sleep efficiency.

***Sociability***: The total number of SMS/WhatsApp message exchanges and GSM/WhatsApp calls which are longer than 1 minute. The number of messages exchanged is the number of "different" contacts with which a person had exchanged messages during a day (midnight is the cut-off). The number of calls > 1 min means the number of times a person had calls longer than 1 minute during a day (midnight is the cut-off). The vertical axis shows the total number of messages exchanged and the number of calls > 1 min, and the horizontal axis shows the duration for which the graph is drawn (see description of CW, PW, and PM below).

***Mobility***: Calculated based on the obfuscated GPS data, to describe the "movement" behavior of the user, considering home as the origin. Home is defined as the GPS location where the person has spent most of the time during the night (between 9 pm and 6 am). The mobility graph is drawn based on the time away from home and the radius of gyration, which are both calculated based on the GPS data. The vertical axis shows the time away from home in hours while the color of the graph indicates the average radius of gyration in kilometers. As shown in Figure 8, these features are calculated by taking the average in following time frames:
- CW – Current Week (7 days before 0000 hours today)
- PW – Past Week (7 days before the current week)
- PM – Past Month (30 days before the past week)

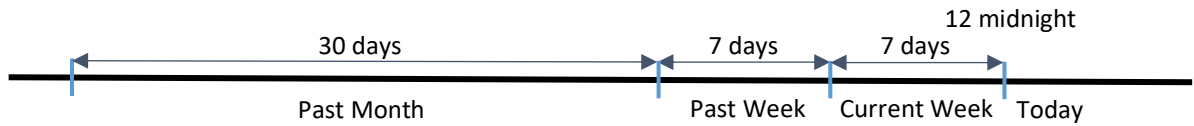

Figure 8: Illustration of the calculation of time periods

## 8.3 Example Scenarios Presented in the Clinician Dashboard

*Scenario 1: as shown in Figure 9a, an exemplary working adult of age 35, goes to the office, actively associates with friends, has kids and is married, generally has 6 hours of good sleep, often takes calls, sends messages to communicate with friends and colleagues, commutes to the office during the weekdays and goes out during the weekend.*

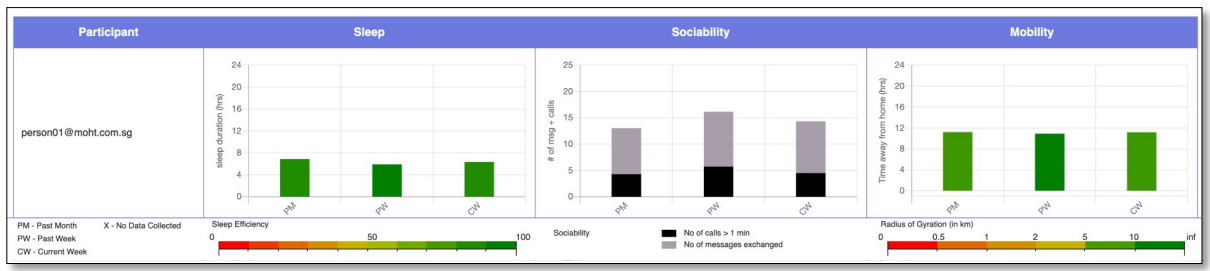

Figure 9a: example scenario 1 for the clinician dashboard

*Observations*: Consistent sleep durations and efficiency. No significant variations in number of calls and messages. Due to the regular behavior, time away from home is consistent and small variations in the radius of gyration occurs due to the places visited during the weekends

*Scenario 2: An exemplary participant with schizophrenia is living with parents and aged in the mid-20s, not going out often and plays video games at home. The participant doesn't have many friends and has few social interactions, sleeps a lot. This participant may be suffering from a relapse.*

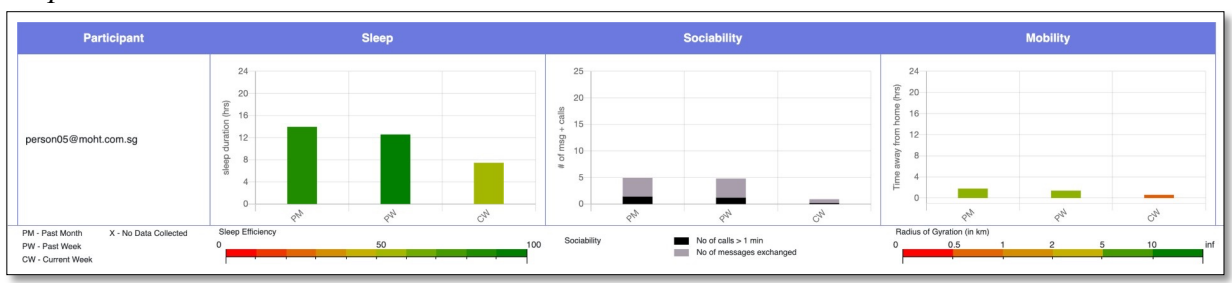

Figure 9b: example scenario 2 for the clinician dashboard

*Observations:* Sleep duration is high compared to a working adult. Similarly, sociability is low compared to a working adult as he doesn't have many friends. Since he preferers to stay at home and play video games, mobility is low compared to a working adult. But because of the possible emerging relapse during the last week, total sleeping hours and sleep efficiency has dropped. Because of pandemic restrictions, this participant may be isolating himself, and sociability and mobility have have dropped significantly.

## 8.4 Data Completion Dashboard

The *data completion dashboard* (see Figure 10) shows the historical completion status of the actual decrypted data from every participant over a customizable period (default is the last 90 days). It is implemented as a component module in the data processing pipeline and thus is updated every time when the data processing pipeline updates.

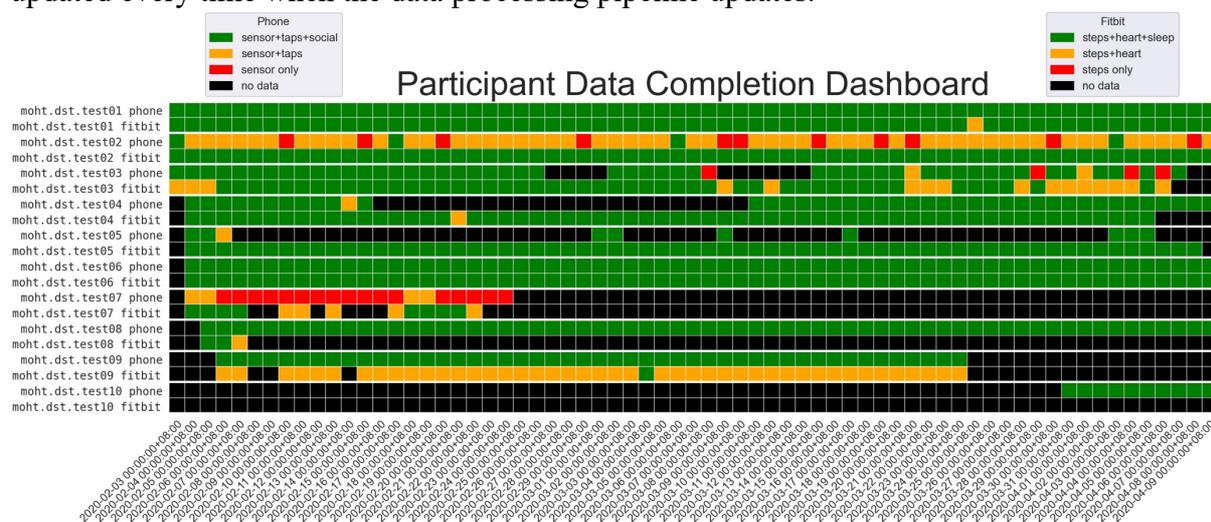

Figure 10: Data completion dashboard showing the completeness of every participant's daily data.

## 8.5 Features Selected in Anomaly Detection Dashboard

We have fitted several univariate time-series models to each of the following 12 *daily* digital phenotyping features; these were motivated by our HOPE-S study:

- sleep mean efficiency - the mean of the sleep efficiency scores during all periods of sleep;
- sleep tot hrs - the total amount of time spent (in hours) asleep;
- # steps - the total number of steps taken throughout the day;
- # walks - the total number of consecutive periods of steps (sampling interval is one minute), which we define as a *walk*;
- steps / min walk - the rate (in steps per minute) during all periods of a walk;
- social # sent - the number of messages and images sent on WhatsApp;
- social # recv - the number of messages and images received on WhatsApp;
- social # contact exch - the number of unique contacts that the participant both sent and received at least one message in WhatsApp;
- # taps - the number of taps in all apps;
- mean intap dur - the mean of the duration of intervals between all screen taps;
- RoG - the radius of gyration as measured by GPS and computed by the *Beiwe* backend;
- light mean lum - the mean recorded lumens by the ambient light sensor.